\documentclass{article}
\usepackage{latexsym}
\usepackage{amssymb}
\usepackage[english]{babel}
\newtheorem{theorem}{Theorem}
\newtheorem{lemma}[theorem]{Lemma}
\begin{document}
\title{The equality problem for infinite words generated by primitive
morphisms}
\author{Juha Honkala\\
Department of Mathematics\\
University of Turku\\
20014 Turku, Finland\\
email: juha.honkala@utu.fi}
\date{}
\maketitle

\begin{abstract}
We study the equality problem for infinite words obtained by iterating
morphisms. In particular, we give a practical algorithm to decide whether or
not two words generated by primitive morphisms are equal.
\end{abstract}

\section{Introduction}
The equality problem for pure morphic words (also called the D0L
$\omega$-equivalence problem) was solved by Culik II and Harju \cite{CH}. A
simpler approach was suggested in \cite{H4}. The algorithms given in
\cite{CH,H4} have very high complexities. No practical algorithm is known in
the general case. For the equality problem of infinite words generated by
polynomially bounded morphisms see \cite{H1}.

The purpose of this paper is to give a practical algorithm for the equality
problem of pure morphic words generated by primitive morphisms. By definition,
a morphism $g:X^{\ast}\longrightarrow X^{\ast}$ is primitive if there is a
positive integer $k$ such that for all $a,b\in X$, $a$ occurs in $g^k(b)$.

To explain our result let $X$ be a finite alphabet and let
$g:X^{\ast}\longrightarrow X^{\ast}$ be a morphism. Let $x\in X$ be a letter
such that $x$ is a prefix of $g(x)$. Then $g^i(x)$ is a prefix of $g^{i+1}(x)$
for all $i\geq 0$. If the length of $g^{i+1}(x)$ is always greater than the
length of $g^i(x)$, we let $g^{\omega}(x)$ be the infinite word which has
prefix $g^i(x)$ for all $i\geq 0$. Infinite words generated in this way by
iterating a morphism are called pure morphic words (or infinite D0L words).

Let now $X$ be an alphabet having $n\geq 2$ letters. Define $A(n)=\lfloor
9n^3\sqrt{n\log n}\rfloor$. Let $g:X^{\ast}\longrightarrow X^{\ast}$ and
$h:X^{\ast}\longrightarrow X^{\ast}$ be primitive morphisms and let $x\in X$
be a letter such that $g^{\omega}(x)$ and $h^{\omega}(x)$ exist. Define
$f_1=g^{2n-2}h^{2n-2}$ and $f_2=h^{2n-2}g^{2n-2}$. We will show that
$g^{\omega}(x)=h^{\omega}(x)$ if and only if $f_1$ and $f_2$ satisfy a balance
condition and one of the words $f_1f_1^{A(n)}(x)$ and $f_2f_1^{A(n)}(x)$ is a
prefix of the other.

The paper is structured as follows. In Section 2 we recall some basic
definitions and earlier results. In particular, we define looping and
loop-free morphisms. Intuitively, a morphism $g:X^{\ast}\longrightarrow
X^{\ast}$ is looping if some power of $g$ generates a periodic word. In
Section 3 we recall the reduction of the equality problem to the comparability
problem. In Section 4 we study balance properties of morphisms. In Section 5
we solve the equality problem for loop-free primitive morphisms by using ideas
and results from \cite{H3}. In Section 6 we solve the equality problem for
looping primitive morphisms. The main result is presented in Section 7.

We assume that the reader is familiar with the basics concerning free monoid
morphisms and their iterations, see \cite{AS,RS1,RS2}. For morphic words
and their applications see also \cite{L}.

\section{Preliminaries}
\subsection{Basic definitions}
We use standard language-theoretic notation and terminology. In particular,
the {\em length} of a word $w$ is denoted by $|w|$. If $w\in X^{\ast}$ is a
nonempty word, then $\mbox{alph}(w)$ is the set of all letters of $X$
occurring in $w$. If $w$ is a nonempty word, then $w^{\omega}$ is the infinite
word
$$w^{\omega}=www\cdots .$$

If $u,v\in X^{\ast}$ are words, $v$ is a {\em factor} of $u$ if there
exist words $u_1,u_2\in X^{\ast}$ such that $u=u_1vu_2$. If, furthermore,
$u_1=\varepsilon$ then $v$ is a {\em prefix} of $u$. If $q$ is a nonnegative
integer and $w$ is a finite word or an infinite word, then $\mbox{Pref}_q(w)$
is the prefix of length $q$ of $w$ and $F_q(w)$ is the set of factors of
length $q$ of $w$. If $|w|< q$, it is understood that $\mbox{Pref}_q(w)=w$. If
$L\subseteq X^{\ast}$, then
$$
\mbox{Pref}_q(L)=\{\mbox{Pref}_q(w)\mid w\in L\}
$$
and
$$
F_q(L)=\bigcup_{w\in L} F_q(w).
$$

Let $u\in X^{\ast}$ be a nonempty word and denote $u=a_1\ldots a_k$ where
$a_i\in X$ for $i=1,\ldots,k$. Then a positive integer $p$ is called a {\em
period} of $u$ if
\begin{equation}
a_{p+i}=a_i \hspace{3mm} \mbox { for } \hspace{3mm} i=1,\ldots,k-p. \label{d1}
\end{equation}
The smallest positive integer $p$ satisfying (\ref{d1}) is called {\em the period} of
$u$ and is denoted by $\mbox{PER}(u)$.

Let $g:X^{\ast}\longrightarrow X^{\ast}$ be a morphism and let $w\in X^{\ast}$
be a word. If $w$ is a prefix of $g(w)$, then $g^n(w)$ is a prefix of
$g^{n+1}(w)$ for all $n\geq 0$. If, furthermore,
$$
\lim_{n\rightarrow \infty} |g^n(w)|=\infty,
$$
then we denote
$$
g^{\omega}(w)=\lim_{n\rightarrow \infty} g^n(w).
$$
In all other cases $g^{\omega}(w)$ is not defined. Hence, if
$g^{\omega}(w)$ exists, it is the unique infinite word which has the
prefix $g^n(w)$ for all $n\geq 0$.

\subsection{Primitive morphisms}

Suppose $g:X^{\ast}\longrightarrow X^{\ast}$ is a morphism. Then $g$ is called
{\em primitive} if there exists a positive integer $k$ such that for all
$x,y\in X$, $y$ occurs in $g^k(x)$. If the alphabet $X$ contains at least two
letters, primitive morphisms are growing morphisms. In general, a morphism
$g:X^{\ast}\longrightarrow X^{\ast}$ is called {\em growing} if for every
$x\in X$ we have
$$
|g^i(x)|\rightarrow \infty \hspace{2mm}\mbox{ when } \hspace{2mm} i\rightarrow
\infty.
$$

If $g:X^{\ast}\longrightarrow X^{\ast}$ is a morphism, define
$$M_g=\max\{|g(x)|\mid x\in X\}$$
and
$$\mbox{CYCLIC}(g)=\{x\in X\mid g(x) \mbox{ contains at least one occurrence
of } x\}.$$

The following lemma gives some basic properties of primitive morphisms.

\begin{lemma}
Let $X$ be an alphabet having $n\geq 2$ letters and let
$g:X^{\ast}\longrightarrow X^{\ast}$ be a primitive morphism. Then
\begin{enumerate}
\item[(i)] $|g^n(x)|\geq M_g$ for all $x\in X$.

\item[(ii)] If $\mbox{\rm CYCLIC}(g)\neq \emptyset$, then $\mbox{\rm alph}(g^{2n-2}(x))=X$
for all $x\in X$.
\end{enumerate}
\end{lemma}
{\em Proof.} Let $x,y\in X$. Because $g$ is primitive, there is an integer $i$
such that $y$ occurs in $g^i(x)$ and $i\leq n-1$. Hence $|g^n(x)|\geq |g(y)|$.
This implies (i).

Assume then that $z\in \mbox{CYCLIC}(g)$. Then $\mbox{alph}(g^{n-1}(z))=X$.
Furthermore, if $x\in X$, then $z$ occurs in $g^{n-1}(x)$. These facts imply
(ii). $\Box$
\medskip

For the proof of the following lemma see \cite{H2}.

\begin{lemma}
Let $X$ be an alphabet with $n\geq 2$ letters and let
$g:X^{\ast}\longrightarrow X^{\ast}$ be a growing morphism. Let $w\in
X^{\ast}$. Then
$$
\mbox{\rm Pref}_2(\{g^i(w)\mid i\geq 0\})=\mbox{\rm Pref}_2(\{g^i(w)\mid
i=0,1,\ldots,3n-2\})
$$
and
$$
F_3(\{g^i(w)\mid i\geq 0\})=F_3(\{g^i(w)\mid i=0,1,\ldots, 2n^2+2n-3\}).
$$
\end{lemma}

\subsection{Looping and loop-free morphisms}

Let $g:X^{\ast}\longrightarrow X^{\ast}$ be a morphism. Then we say that $g$
is {\em looping} if there exist a positive integer $k$, a letter $x\in X$ and
a nonempty word $u$ such that $(g^k)^{\omega}(x)$ exists and
$$(g^k)^{\omega}(x)=u^{\omega}.$$
If $g$ is not looping, we say that $g$ is {\em loop-free}.

Loop-free morphisms avoid small periods in the sense of the following lemma
\cite{H3}.

\begin{lemma}
Let $X$ be an alphabet having $n$ letters and let $g:X^{\ast}\longrightarrow
X^{\ast}$ be a morphism. If $g$ is growing and loop-free, then
$$
\mbox{\rm PER}(g^d(x))\geq \frac{1}{2M_g^n}|g^d(x)|
$$
whenever $x\in X$ and $d$ is a positive integer.
\end{lemma}

For the proof of the next lemma see \cite{H5}.

\begin{lemma}
Let $X$ be an alphabet having $n$ letters and let $g:X^{\ast}\longrightarrow
X^{\ast}$ be a primitive morphism. Assume that $x\in X$, $k$ is a positive
integer and
$$(g^k)^{\omega}(x)=u^{\omega},$$
where $u\in X^{\ast}$ is a primitive word. Then
$$|g^{2n}(z)|\geq 2|u|$$
for all $z\in X$.
\end{lemma}

\section{Properties of infinite words}
In this section we recall some results concerning infinite words generated by
morphisms.

Let $g:X^{\ast}\longrightarrow X^{\ast}$ be a morphism and let $x\in X$ be a
letter such that $g^{\omega}(x)$ exists. If $w$ is a nonempty prefix of
$g^{\omega}(x)$, then $w$ is a proper prefix of $g(w)$.

\begin{lemma}
Let $g_i:X^{\ast}\longrightarrow X^{\ast}$, $i=1,2$, be morphisms and let
$x\in X$ be a letter such that $g_i^{\omega}(x)$ exists for $i=1,2$. Suppose
$$g_i^{\omega}(x)=wb_i\ldots$$
for $i=1,2$, where $w\in X^{\ast}$, $b_i\in X$ and $b_1\neq b_2$. If
$i_1,\ldots,i_k\in \{1,2\}$, then either $g_{i_k}\ldots g_{i_1}(x)$ is a
prefix of $w$ or has prefix $wb_{i_k}$.
\end{lemma}
{\em Proof.} Consider the word $g_{i_k}\ldots g_{i_1}(x)$ and assume that the
claim holds for $g_{i_{k-1}}\ldots g_{i_1}(x)$. If
$g_{i_{k-1}}\ldots g_{i_1}(x)$ is a prefix of $w$, the claim holds. Assume $w$
is prefix of $g_{i_{k-1}}\ldots g_{i_1}(x)$. Then $g_{i_k}(w)$ is a prefix of
$g_{i_k}\ldots g_{i_1}(x)$. The claim follows because $g_{i_k}(w)$ is a prefix
of $g_{i_k}^{\omega}(x)$ which is longer than $w$. $\Box$

\begin{lemma}
Let $g:X^{\ast}\longrightarrow X^{\ast}$ and $h:X^{\ast}\longrightarrow
X^{\ast}$ be morphisms and let $x\in X$ be a letter such that $g^{\omega}(x)$,
$h^{\omega}(x)$, $(gh)^{\omega}(x)$ and $(hg)^{\omega}(x)$ exist. Then
\begin{equation}
g^{\omega}(x)=h^{\omega}(x)
\label{r1}
\end{equation}
if and only if
\begin{equation}
(gh)^{\omega}(x)=(hg)^{\omega}(x).
\label{r2}
\end{equation}
\end{lemma}
{\bf Proof.} Assume first that (\ref{r1}) holds. If $u\in
\mbox{Pref}(g^{\omega}(x))$, then $gh(u)\in \mbox{Pref}(g^{\omega}(x))$. Hence
$$(gh)^i(x)\in \mbox{Pref}(g^{\omega}(x))$$
for all $i\geq 0$. Therefore
$$(gh)^{\omega}(x)=g^{\omega}(x).$$
Similarly
$$(hg)^{\omega}(x)=g^{\omega}(x).$$
Hence (\ref{r2}) holds.

Assume then that (\ref{r1}) does not hold. Let
$$g^{\omega}(x)=wa\cdots, \hspace{3mm} h^{\omega}(x)=wb\cdots,$$
where $w\in X^{\ast}$, $a,b\in X$ and $a\neq b$. Then for large values of $i$
by Lemma 5, $wa$ is a prefix of $(gh)^i(x)$ and $wb$ is a prefix of
$(hg)^i(x)$. Hence (\ref{r2}) does not hold. $\Box$
\medskip

Let $g:X^{\ast}\longrightarrow X^{\ast}$ and $h:X^{\ast}\longrightarrow X^{\ast}$ be
morphisms. Then the set \linebreak $\mbox{COMP}(g,h)$ is defined by
\begin{eqnarray*}
\mbox{COMP}(g,h)&=&\{w\in X^{\ast}\mid \mbox{there exist } u_1,u_2\in
X^{\ast} \mbox{ such that } \\
& & g(w)u_1=h(w)u_2\}.
\end{eqnarray*}

\begin{lemma}
Let $g:X^{\ast}\longrightarrow X^{\ast}$ and $h:X^{\ast}\longrightarrow
X^{\ast}$ be morphisms and let $x\in X$ be a letter such that $g^{\omega}(x)$
and $h^{\omega}(x)$ exist. Then
\begin{equation}
g^{\omega}(x)=h^{\omega}(x)
\label{r3}
\end{equation}
if and only if
\begin{equation}
\mbox{\rm Pref}(g^{\omega}(x))\subseteq \mbox{\rm COMP}(g,h).
\label{r4}
\end{equation}
\end{lemma}
{\em Proof.} Assume first that (\ref{r3}) holds. Assume that $w$ is a prefix
of $g^{\omega}(x)=h^{\omega}(x)$. Then $g(w)$ is a prefix of $g^{\omega}(x)$
and $h(w)$ is a prefix of $h^{\omega}(x)$. Hence $w\in \mbox{COMP}(g,h)$.

Assume then that (\ref{r3}) does not hold. Let
$$g^{\omega}(x)=wa\cdots, \hspace{3mm} h^{\omega}(x)=wb\cdots,$$
where $w\in X^{\ast}$, $a,b\in X$ and $a\neq b$. Then $g(w)$ has prefix $wa$
and $h(w)$ has prefix $wb$. Because $a\neq b$, the word $w$ is not in
$\mbox{COMP}(g,h)$ which shows that (\ref{r4}) does not hold. $\Box$

\section{Balance properties of morphisms}
Let $g:X^{\ast}\longrightarrow X^{\ast}$ and $h:X^{\ast}\longrightarrow
X^{\ast}$ be morphisms. Define
$$\mbox{BAL}(g,h)=\max\{||gg^k(x)|-|hg^k(x)||\mid x\in X, k\geq 0\},$$
where the right-hand side is a nonnegative integer or $\infty$.

For the proof of the following lemma see \cite{H2}.

\begin{lemma}
Let $X$ be an alphabet having $n\geq 2$ letters and let
$g:X^{\ast}\longrightarrow X^{\ast}$ and $h:X^{\ast}\longrightarrow X^{\ast}$
be morphisms. Define $M=\max\{M_g,M_h\}$. If
$$\mbox{\rm BAL}(g,h)<\infty,$$
then
$$\mbox{\rm BAL}(g,h) \leq  M^{2n-1}\exp(n^2(1+\sqrt{6n\log n})).$$
\end{lemma}

Next we recall a very important result due to Culik II and Harju \cite{CH}.
Let $g:X^{\ast}\longrightarrow X^{\ast}$ be a morphism and let $a\in X$ be a
letter. Define $X_1=X-\{a\}$. Then the triple $(X,g,a)$ is called a {\em
1-system} if
\begin{enumerate}
\item[(i)]
$g(a)\in aX_1^{\ast}$,
\item[(ii)]
$g(x)\in X_1^{\ast}$ whenever $x\in X_1$,
\item[(iii)]
$\{g^i(a)\mid i\geq 0\}$ is infinite,
\item[(iv)]
if $x\in X_1$, then $x$ occurs infinitely many times in $g^{\omega}(a)$.
\end{enumerate}
A 1-system $(X,g,a)$ is called {\em 1-simple} if the restriction of $g$ to
$X_1$ is primitive.

For the proof of the following result see \cite{CH}.

\begin{theorem}
Let $(X,g_i,a)$, $i=1,2$, be 1-systems such that
$g_1^{\omega}(a)=g_2^{\omega}(a)$. If $(X,g_1g_2,a)$ and $(X,g_2g_1,a)$ are
1-simple then there exists a positive integer $K$ such that
$$||g_1g_2(w)|-|g_2g_1(w)||\leq K$$
whenever $w$ is a prefix of $g_1^{\omega}(a)$.
\end{theorem}

In what follows we need the following consequence of Theorem 9.

\begin{lemma}
Let $X_1$ be a finite alphabet.
Let $g_1:X_1^{\ast}\longrightarrow X_1^{\ast}$ and $g_2:X_1^{\ast}\longrightarrow
X_1^{\ast}$ be primitive morphisms and let $x\in X_1$ be a letter such that
$g_1^{\omega}(x)$ and $g_2^{\omega}(x)$ exist and are equal. If $g_1g_2$ and
$g_2g_1$ are primitive then there exists a positive integer $K$ such that
$$||g_1g_2(w)|-|g_2g_1(w)||\leq K$$
whenever $w$ is a factor of $g_1^{\omega}(x)$.
\end{lemma}
{\em Proof.} It is enough to prove the claim for the prefixes of
$g_1^{\omega}(x)$. This claim is a consequence of Theorem 9. Indeed, choose a
new letter $a\not\in X_1$ and define $X=X_1\cup \{a\}$. Assume that
$g_i(x)=xu_i$ ($i=1,2$) and extend $g_i$ by $g_i(a)=au_i$ ($i=1,2$). Then
$(X,g_i,a)$, $i=1,2$, are 1-systems and $g_1^{\omega}(a)=g_2^{\omega}(a)$.
Furthermore, $(X,g_1g_2,a)$ and $(X,g_2g_1,a)$ are 1-simple. Hence Theorem
9 is applicable. $\Box$

\section{The equality problem for loop-free primitive morphisms}
\subsection{The comparability problem}
In this subsection we first recall a result from \cite{H3}.

Let $X$ be an alphabet having $n\geq 2$ letters and let
$g:X^{\ast}\longrightarrow X^{\ast}$ and $h:X^{\ast}\longrightarrow X^{\ast}$
be growing morphisms. Define the mapping $\beta: X^{\ast}\longrightarrow
{\mathbb Z}$ by
$$
\beta(w)=|g(w)|-|h(w)|, \hspace{3mm} w\in X^{\ast}.
$$

Assume that $t\geq 2$ is a fixed integer. Assume that $d$ is a positive
integer such that we have
\begin{equation}
\mbox{PER}(g^d(z))\geq \frac{1}{t} |g^d(z)| \hspace{3mm} \mbox{ for all }
\hspace{3mm} z\in X. \label{m1}
\end{equation}
Define
$$e=d+2n-1$$
and
$$
B=\max_{z\in X}|\beta(g^e(z))|.
$$

Further, assume that $m$ is a positive integer such that
\begin{equation}
|g^d(z)|>2tm \hspace{2mm} \mbox{ for all } \hspace{2mm} z\in X
\label{m2}
\end{equation}
and
\begin{equation}
B\leq \frac{m}{n^2+1}.
\label{m3}
\end{equation}

Define
$$
L_1=\mbox{COMP}(hg^e,gg^e).
$$
For the proof of the next lemma see \cite[Lemma 9]{H3}.

\begin{lemma}
Let $w\in X^{\ast}$. Then
$$w\in L_1$$
if and only if
$$\mbox{\rm Pref}_2(w)\in L_1 \mbox{ and } F_3(w)\subseteq F_3(L_1).$$
\end{lemma}

Now we use Lemmas 2 and 11 to show that for a word $w\in X^{\ast}$ we can
decide the inclusion
$$\{g^i(w)\mid i\geq 0\}\subseteq \mbox{COMP}(g,h)$$
by checking whether or not $g^i(w)\in \mbox{COMP}(g,h)$ for a certain number
of initial values of $i$.

\begin{lemma}
Let $w\in X^{\ast}$. Then
$$\{g^i(w)\mid i\geq 0\}\subseteq \mbox{\rm COMP}(g,h)$$
if and only if
$$\{g^i(w)\mid i=0,1,\ldots,e+2n^2+2n-3\}\subseteq \mbox{\rm COMP}(g,h).$$
\end{lemma}
{\em Proof.} To prove the nontrivial part assume that $g^i(w)\in
\mbox{COMP}(g,h)$ for $i=0,1,\ldots,e+2n^2+2n-3$. Then $g^i(w)\in L_1$
for $i=0,1,\ldots,2n^2+2n-3$. Hence
\begin{equation}
\mbox{Pref}_2(g^i(w))\in L_1 \mbox{ and } F_3(g^i(w))\subseteq F_3(L_1)
\label{m4}
\end{equation}
for $i=0,1,\ldots,2n^2+2n-3$. Now Lemma 2 implies that (\ref{m4}) holds
for all $i\geq 0$. Therefore Lemma 11 implies that $g^i(w)\in L_1$ for all
$i\geq 0$. In other words $g^{e+i}(w)\in \mbox{COMP}(g,h)$ for all $i\geq 0$.
Hence $g^i(w)\in \mbox{COMP}(g,h)$ for all $i\geq 0$. $\Box$

\subsection{The equality problem}
In this subsection we solve the equality problem for pure morphic words
generated by loop-free primitive morphisms under certain additional
assumptions.

\begin{lemma}
Let $X$ be an alphabet having $n\geq 2$ letters. Define $A(n)=$ \linebreak $\lfloor
9n^3\sqrt{n\log n}\rfloor$. Let
$g:X^{\ast}\longrightarrow X^{\ast}$ and $h:X^{\ast}\longrightarrow X^{\ast}$
be primitive morphisms. Let $x\in X$ be a letter such that $g^{\omega}(x)$ and
$h^{\omega}(x)$ exist. Assume that $g$ is loop-free,
$\mbox{\rm BAL}(g,h)<\infty$ and $M_h\leq M_g^2$. Then
\begin{equation}
g^{\omega}(x)=h^{\omega}(x)
\label{m5}
\end{equation}
if and only if
\begin{equation}
g^{A(n)}(x)\in \mbox{\rm COMP}(g,h).
\label{m6}
\end{equation}
\end{lemma}
{\em Proof.} First, if (\ref{m5}) holds, Lemma 7 implies (\ref{m6}).

Assume then that (\ref{m6}) holds. Define the integers $t,m,d,e$ and $B$ as
follows. First, define
$$t=2M_g^n$$
and
$$m=\lceil (n^2+1)M_g^{2(2n-1)}\exp(n^2(1+\sqrt{6n\log n}))\rceil .$$
Let $q$ be a real number such that
$$2tm=M_g^q.$$
Then define
$$d=(\lfloor q \rfloor +1)n, \hspace{3mm} e=d+2n-1.$$
Finally, define
$$B=\max_{z\in X}|\beta(g^e(z))|.$$
Then
\begin{equation}
e+2n^2+2n-2 < 9n^3\sqrt{n\log n}. \label{f4}
\end{equation}

Now we are in a position to apply Lemma 12. First, $g$ and $h$ are growing
because $g$ and $h$ are primitive and $\mbox{card}(X)\geq 2$. Because $g$ is
loop-free, Lemma 3 implies (\ref{m1}). By Lemma 1 we have
$$|g^d(z)|\geq M_g^{\lfloor q \rfloor +1}>2tm$$
for all $z\in X$. Hence (\ref{m2}) holds. By Lemma 8 also (\ref{m3}) holds.
Observe that the assumption $M_h\leq M_g^2$ implies that
$M=\max\{M_g,M_h\}\leq M_g^2$.
Consequently the assumptions needed for Lemma 12 hold.
By (\ref{f4}) Lemma 12 implies that $g^i(x)\in \mbox{COMP}(g,h)$ for all $i\geq
0$. Now Lemma 7 implies (\ref{m5}). $\Box$

\section{The equality problem for looping primitive morphisms}
In this section we solve the equality problem for pure morphic words generated
by looping primitive morphisms under some additional assumptions.
\begin{lemma}
Let $X$ be an alphabet having $n\geq 2$ letters. Let
$g:X^{\ast}\longrightarrow X^{\ast}$ and $h:X^{\ast}\longrightarrow X^{\ast}$
be primitive morphisms. Let $x\in X$ be a letter such that $g^{\omega}(x)$ and
$h^{\omega}(x)$ exist. Assume that $g$ is looping and
$\mbox{\rm BAL}(g,h)<\infty$. Then
\begin{equation}
g^{\omega}(x)=h^{\omega}(x)
\label{l1}
\end{equation}
if and only if
\begin{equation}
g^{2n}(x)\in \mbox{\rm COMP}(g,h).
\label{l2}
\end{equation}
\end{lemma}
{\em Proof.} Because $g$ is looping there exist a positive integer $k$, a
letter $z\in X$ and a nonempty primitive word $v$ such that
$$(g^k)^{\omega}(z)=v^{\omega}.$$
Because $g$ is primitive, each letter of $X$ occurs in $v$. Because $g^k(v)\in
v^{\ast}$ it follows that $g^{ki}(x)$ is a factor of $v^{\omega}$ for all
$i\geq 0$.  This implies that there is a conjugate $u$ of $v$ such that for
infinitely many values of $i$, the word $g^{ki}(x)$ is a prefix of
$u^{\omega}$. Because $g^{ki}(x)$ is a prefix of $g^{\omega}(x)$ for all
$i\geq 0$, we have
\begin{equation}
g^{\omega}(x)=u^{\omega}.
\label{l3}
\end{equation}
By Lemma 4 we have
\begin{equation}
|g^{2n}(x)|\geq 2|u|.
\label{l4}
\end{equation}
Because $\mbox{BAL}(g,h)<\infty$, we have $|g(u)|=|h(u)|$.

Assume now that (\ref{l2}) holds. By (\ref{l3}) and (\ref{l4}), the word $u$
is a prefix of $g^{2n}(x)$. Hence $u\in \mbox{COMP}(g,h)$, which implies that
$g(u)=h(u)$. Therefore
$g^i(x)\in \mbox{COMP}(g,h)$ for all $i\geq 0$. This implies (\ref{l1}) by
Lemma 7.

Conversely, if (\ref{l1}) holds, Lemma 7 implies (\ref{l2}). $\Box$

\section{The equality problem for primitive morphisms}
In the previous sections we have studied the equality
$g^{\omega}(x)=h^{\omega}(x)$ with the assumption that
$\mbox{BAL}(g,h)<\infty$. It remains to give a necessary and sufficient
condition for the equality $g^{\omega}(x)=h^{\omega}(x)$ without the
assumption that $\mbox{BAL}(g,h)<\infty$.

\begin{theorem}
Let $X$ be an alphabet having $n\geq 2$ letters. Define $A(n)=\lfloor
9n^3\sqrt{n\log n}\rfloor$. Let $g:X^{\ast}\longrightarrow X^{\ast}$ and
$h:X^{\ast}\longrightarrow X^{\ast}$ be primitive morphisms and let $x\in X$
be a letter such that $g^{\omega}(x)$ and $h^{\omega}(x)$ exist. Define
$f_1=g^{2n-2}h^{2n-2}$ and $f_2=h^{2n-2}g^{2n-2}$.
Then
\begin{equation}
g^{\omega}(x)=h^{\omega}(x)
\label{t1}
\end{equation}
if and only if
\begin{equation}
\mbox{\rm BAL}(f_1,f_2)<\infty
\label{t2}
\end{equation}
and
\begin{equation}
f_1^{A(n)}(x)\in \mbox{\rm COMP}(f_1,f_2).
\label{t3}
\end{equation}
\end{theorem}
{\bf Proof.} By Lemma 1,
\begin{equation}
\mbox{alph}(g^{2n-2}(z))=\mbox{alph}(h^{2n-2}(z))=X
\label{t4}
\end{equation}
for all $z\in X$. Hence $f_1$ and $f_2$ are primitive morphisms. In
particular, $f_1$ and $f_2$ are growing morphisms and $f_1^{\omega}(x)$ and
$f_2^{\omega}(x)$ exist. If $y,z\in X$, then by (\ref{t4}) we have
$$|f_2(y)|=|h^{2n-2}g^{2n-2}(y)|\leq |h^{2n-2}g^{2n-2}h^{2n-2}(z)|\leq
|f_1^2(z)|.$$
Hence
$$M_{f_2}\leq M_{f_1^2}\leq M_{f_1}^2.$$

Now suppose that (\ref{t1}) holds. Then Lemma 6 implies that
\begin{equation}
f_1^{\omega}(x)=f_2^{\omega}(x).
\label{t5}
\end{equation}
Now Lemma 10 implies (\ref{t2}) and Lemma 7 implies (\ref{t3}).

Assume then that (\ref{t2}) and (\ref{t3}) hold. If $f_1$ is loop-free (resp.
looping), then Lemma 13 (resp. Lemma 14) implies that (\ref{t5}) holds. Then
we have (\ref{t1}) by Lemma 6. $\Box$
\medskip

For a method to decide the condition (\ref{t2}) see \cite[p. 512]{H2}, where
it is explained how one can compute polynomials $P_1(z),\ldots,P_n(z)$
depending on $f_1$ and $f_2$ such that (\ref{t2}) holds if and only if for
each $j$ there is a positive integer $p\leq \exp(\sqrt{6n\log n})$ such that
$P_j(z)$ divides $1-z^p$. Hence the degrees of the polynomials depend only on
the cardinality of the alphabet.

\end{document}